# Measurement and Shaping of Perfect Optical Vortex via Cross Phase

**Chen Wang (王 琛)[1,2], Yuan Ren (任 元)[1,2]\*, Tong Liu(刘 通)[1,2], Zhengliang Liu(刘政良)[1,2], Song Qiu(邱 松)[1,2], Zhimeng Li(李智猛)[1,2], You Ding(丁 友)[1,2], and Hao Wu(吴 昊)[1,2]**

[1]*Department of Aerospace Science and Technology, Space Engineering University, Beijing 101416, China*

[2]*Lab of Quantum Detection & Awareness, Space Engineering University, Beijing 101416, China*

*\*Corresponding author: renyuan_823@aliyun.com;*

We investigate a method for the measurement and shaping of a perfect optical vortex (POV) by using the cross-phase for the first time. Experimentally, we propose a method to measure the topological charges +3 and ±5 of POVs with the low-order cross-phase (LOCP) at the Fourier plane; we realize shaping intensity distributions of POVs at the Fourier plane with the high-order cross-phase (HOCP) and show the asymmetric modulation ability of the HOCP for an optical field. Furthermore, we discuss the influence of parameters on the optical field in the shaping process. On this basis, we propose a secondary energy control method that combines the characteristics of the POV, which ensures flexible control of the shape while reducing the impact on the orbital angular momentum of the original POV. This work provides an alternative method to measure and shape a POV, which facilitates applications in optical micro-manipulation.

doi:10.3788/COLXXXXXX.XXXXXX.

An optical vortex with the phase factor exp(i$m\phi$), carry orbital angular momentum (OAM) of $m\hbar$ per photon, where $m$ denotes the topological charges (TCs), $\phi$ denotes the azimuthal angle and $\hbar$ is the Planck constant. As early as 1992, Allen showed that photons could carry OAM, which has aroused widespread concern among researchers[1]. Since then, optical vortices have been utilized in a plethora of applications in the field of optical micro-manipulation[2], quantum entanglement[3], and rotation speed measurement via the optical rotation Doppler effect[4].

As a kind of optical vortices, perfect optical vortices (POVs) have received intense interest since the radius of a POV is independent of TCs. It's vital that we can measure the TCs, and adjust the shape of a POV, which is a critical issue in the field of optical micro-manipulation and optical communication[5]. However, POVs only appear at the Fourier plane of a spatial light modulator (SLM), which makes it difficult to measure the TCs with conventional interference or diffraction methods[6, 7]. Besides, lots of methods have been proposed to achieve the shaping[8-11] or singularity manipulation of optical vortices[12, 13]. However, few methods can achieve these two functions simultaneously. Fortunately, we have an opportunity to achieve these goals at the same time by adopting the cross-phase (CP).

In 2019, the CP, a new kind of phase structure, has been involved in Laguerre-Gauss (LG) beams that open up a new horizon for generation and measurement of optical vortices[6, 14]. Recently, we investigated a generation and measurement method of high-order optical vortices via the CP, which has been experimentally achieved[15]. In August 2020, we proposed a new type of CP, which can be employed to modulate optical vortices to implement shaping and multi-singularity manipulation simultaneously in the far-field[16]. In October, inspired by the CP, we proposed a new kind of optical vortex called the Hermite–Gaussian-like optical vortex[17].

The form of the CP $\psi_0(x, y)$ in Cartesian coordinates $(x, y)$ is

$$\psi_0(x, y) = u(x^p \cos\theta - y^q \sin\theta)(x^p \sin\theta + y^q \cos\theta) \quad (1)$$

where the coefficient $u$ controls the conversion rate, the azimuth factor $\theta$ characterizes the rotation angle of converted beams in one certain plane. The order $p$ and $q$ are positive integers. When two orders are both equal to 1 (the sum is 2), we call it low-order CP (LOCP), as shown in Fig. 1(a); when the sum of these two orders is greater than 2, we call it high-order CP (HOCP), as shown in Fig. 1(b)-(c).

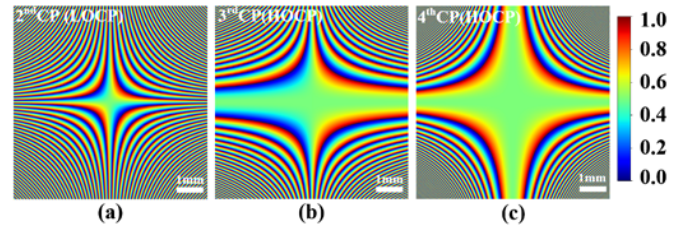

Fig. 1 Phase distributions of the CP. (a) LOCP. (b) HOCP. The order of the HOCP is 3. (c) HOCP. The order of the HOCP is 4.

The LOCP can be engaged in the measurement of TCs; the HOCP could be used to implement shaping, even singularity manipulation of a POV. It is noteworthy that eq.(1) could be simplified to $\psi(x, y) = ux^p y^q$ when $\theta = 0$, and we only take this typical situation into count.

In this Letter, we apply the CP to POVs for the first time. On the one hand, we propose a method to realize the measurement of TCs at the Fourier plane of an SLM via the LOCP; on the other hand, we realize shaping intensity distributions to further improve the performance of POVs via the HOCP. We also discuss the influence of parameter adjustment on beam shaping via the HOCP. On this basis, we propose a secondary energy control method that combines the characteristics of the POV, which ensures flexible control of the shape while reducing the impact on the orbital angular momentum of the original POV.

The CP is based on astigmatism, which shares a similar principle with the astigmatic mode converter. However, the latter has high requirements on the accuracy, such as the relative position of the cylindrical lenses and the oblique incidence angle, etc. In contrast, the HOCP is more conducive to precise particle manipulation and greatly evade the harsh requirements of the astigmatic mode converter with the help of an SLM, which is of great value in the field of optical micro-manipulation [17].

Without loss of generality, may the idea form of a POV:

$$E_1(r,\phi) \equiv \delta(r-R)\exp(im\phi) \quad (2)$$

where $\delta(r-R)$ is the Dirac function, $m$ are TCs, $\phi$ denotes the azimuthal angle, and $\hbar$ is the Planck constant. We can choose a suitable function to replace the Dirac function to generate a POV approximately. The approximation of Eqs.(2) can be obtained by performing Fourier transform on the idea Bessel beam. However, the energy of the idea Bessel beam is infinite and cannot be generated experimentally. We can get an approximate Bessel beam in the experiment, which is the Bessel-Gauss beam:

$$E_2(r,\phi) \equiv J_n(k_r r)\exp(-r^2/\omega^2)\exp(im\phi) \quad (3)$$

where $J_n$ is the n-order Bessel function of the first kind, $k_r$ is the radial wave number, and $\omega$ is the waist radius of the incident beam. After the Fourier transform of Eqs.(3), the expression of a POV can be obtained under the experimental conditions:

$$E_3(r,\phi) \equiv i^{n-1}\frac{\omega}{\omega_0}I_n\left(\frac{2Rr}{\omega_0^2}\right)\exp\left(\frac{-r^2-R^2}{\omega_0^2}\right)\exp(im\phi) \quad (4)$$

where $\omega_0$ is the waist radius of the Gaussian beam at the back focal plane of the lens, $I_n$ is the n-order modified Bessel function of the first kind, and $R$ denotes the radius of the POV.

In this Letter, the POV is generated according to the method reported by Vaity, etc.[5, 18]. The process is as follows: an optical vortex illuminates an axicon, and then the transmitted light becomes a Bessel–Gauss beam; after the Fourier transform of the Bessel–Gauss beam, the POV is generated. The transmission function of the axicon is

$$t(r) = \begin{cases} \exp[-ik(n-1)r\alpha], & (r \leq R_0) \\ 0, & (r > R_0) \end{cases} \quad (5)$$

where $\alpha$ and $n$ are the cone angle and refraction index, respectively. In order to achieve the modulation of a POV via the CP, the generation process of the phase mask is shown in Fig. 2, which is designed by axicon phase multiplying the spiral phase, the CP, and adding a blazed grating. Consequently, the generated is written into the SLM, as shown in Fig. 2(e). This process is equivalent to the interference record of the holography.

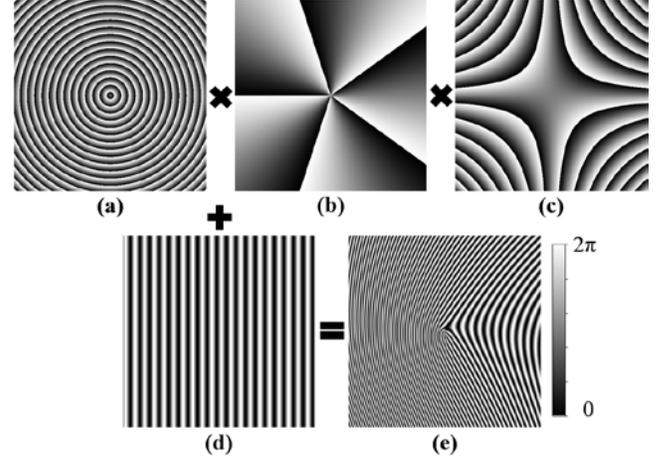

Fig. 2 The generation process of the phase mask. (a) The phase of the axicon. (b) The phase of the optical vortex. (c) The phase of the CP. (d) The phase of the blazed grating. (e) The generated phase mask.

The experimental setup employed to implement both measurement and shaping is shown in Fig. 3. The laser delivers a collimated Gaussian beam with wavelengths of 632.8nm after a linear polarizer (LP), a half-wave plate (HWP), and a telescope consists of two lenses (L1, L2) are used for collimation. The combination of the LP and the HWP is served to rotate the laser polarization state along the long display axis of SLM and adjust the power of incident light on SLM. The SLM (HDSLM80R) precisely modulates the incident light via loading a hologram, and then the aperture (AP) is used to select the first diffraction order of the beam to avoid other stray light. A CCD camera registers the intensity pattern (NEWPORT LBP2) placed at the back focal plane of L3.

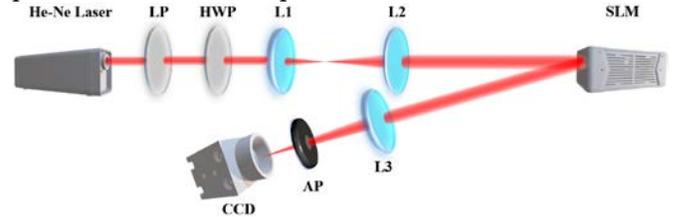

Fig. 3 Experimental setup for the measurement and shaping of POVs via the HOCP. LP: linear polarizer. HWP: half-wave plate. L: lens. SLM: spatial light modulator. AP: aperture. CCD: charge-coupled device.

Firstly, we would like to propose a method to measure POV via the LOCP. The intensity distribution after the measurement is not strictly a Hermit-Gauss distribution but a spindle distribution, as shown in Fig. 4(a). There is a "needle" at each end of the "spindle." We need to calculate how many "black yarns" are between the two needles to confirm the TCs of a POV. Take a POV with TCs of 5 as an example, as shown in the first row of Fig. 4(b), the white dotted lines denote the positions of two "needles," and solid lines denote the positions of "black yarns" in the partially enlarged view of the center. We can easily get that there are 5 "black yarns" between the two "needles." In addition to determining the value of TCs, this method can also determine the symbol, as shown in the second row of Fig. 4(b). If the "needle" at the bottom right is down, the sign of the TC is positive; if the "needle" at the bottom right is up, the sign of the TC is negative. In the partially enlarged views of the center, the white arrows mark the position of the "needle" at the bottom right. The phase distributions and the experimental results corresponding to (b) are shown in Fig. 4(c) and Fig. 4(d), respectively. Compared with the far-field, this method can achieve the measurement of TCs in a short distance, saving space and optical components. Furthermore, this method realizes the in-situ measurement of POVs at the Fourier plane, which is meaningful in optical micro-manipulation.

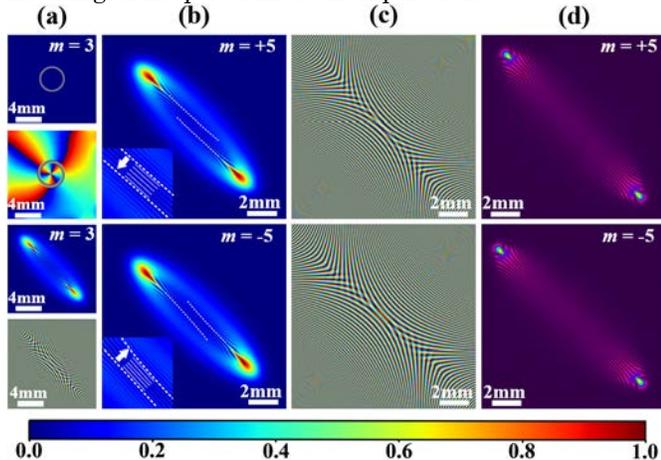

Fig. 4 Measurement of a POV with the LOCP. (a) The first row is the intensity distribution of the original POV without the LOCP at the Fourier plane, and the TCs here are 3. The second row is the phase distribution corresponding to the first row. The third row is the intensity distribution of the original POV with the LOCP at the Fourier plane. The fourth row is the phase distribution corresponding to the third row. (b) Intensity distributions of POVs measured by the LOCP. The top right corner of the figure identifies the TCs of POVs before measurement. The lower-left corner is a partially enlarged view of the center. (c) Phase distributions corresponding to (b). (d) Experimental results corresponding to (b).

Secondly, we would like to introduce the polygonal shaping ability of the HOCP. The HOCP can shape the POV into any symmetrical polygon theoretically. Interestingly, the number of sides of the shaped polygon is strictly equal to the order of the HOCP and not related to TCs. As illustrated in Fig. 5, the TCs of POVs are all 3, and the order of the HOCP is 0 (without the HOCP), 3, 4, 5, correspondingly. Consequently, the corresponding intensity distributions are circle, triangle, quadrangle, and pentagon at the Fourier plane. Because manipulated particles move in the direction of energy flow, and changes in the intensity distribution can affect the distribution of OAM density, we can control the trajectory of the manipulated particles precisely by altering the order of the HOCP flexibly according to actual needs.

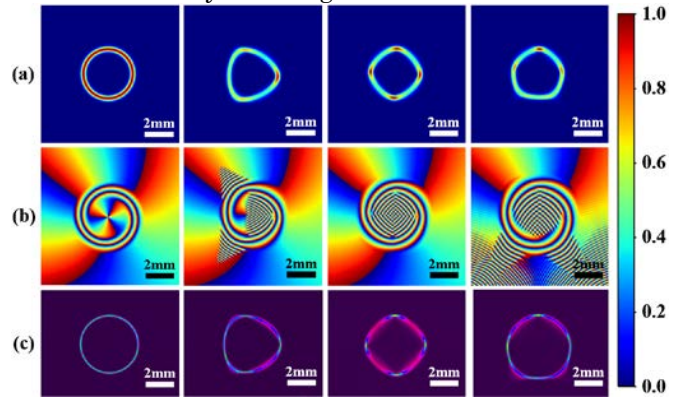

Fig. 5 Polygonal shaping of POVs via the HOCP at the Fourier plane and the TCs of POVs are all 3. (a) Simulated polygonal intensity distributions of POVs shaped by the HOCP. The order of the HOCP is 0 (without the HOCP), 3, 4, 5, correspondingly. (b) Simulated phase distributions corresponding to (a). (c) Experimental intensity distributions corresponding to (a).

Notably, index combinations could affect the symmetry of optical field distributions: the larger the absolute value of the difference between $p$ and $q$ in $\psi(x, y) = ux^p y^q$, the smaller the symmetry of the optical field, which is demonstrated in Fig. 6(a) and Fig. 6(c). The phenomenon motioned above could be explained that the difference between $m$ and $n$ in the HOCP yields asymmetric phase distributions, as shown in Fig. 6 (b) and Fig. 6 (d), performing the asymmetric shaping of POVs. The asymmetric modulation of the optical field greatly expands the range of the modulation of POVs and provides more flexibility in optical micro-manipulation.

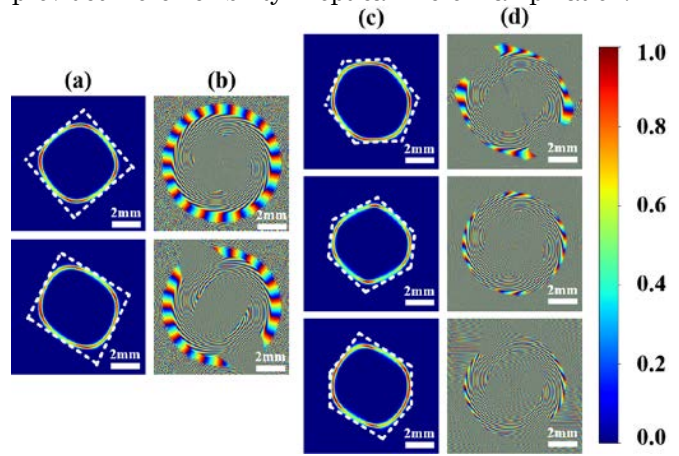

Fig. 6 Simulated results of polygonal POVs modulated by the HOCP. (a) Simulated intensity distributions of pentagonal POVs and the order of the HOCP is 4. The combinations of $p$ and $q$ are 2-2 and 1-3, respectively. (b) Simulated phase distributions of the HOCP corresponding to (a). (c) Simulated intensity distributions of hexagonal POVs and the order of the HOCP is 6. The combinations of $m$ and $n$ are 3-3, 2-4, and 1-5, respectively. (d) Simulated phase distributions of the HOCP corresponding to (c).

Finally, we would like to discuss the influence of parameters on the distribution of the optical field. As we all know, the radius and the TC of a POV are decoupled. The POV modulated by the HOCP still retains this feature, as shown in Fig. 7. As the TC changes from 1 to 4, the modulated POV still maintains its original shape and size, but the energy of the beam gradually converges to the 4 vertices. As we all know, when a polygonal optical vortex is used to confine particles, the particles will move with varying acceleration and deceleration along with the intensity distribution. When the movement reaches the vertex, the particle will accelerate. When moving away from the vertex, the particles will slow down. In the shaping process, the control of the energy at the vertex can further accurately control the particle acceleration and realize more complex particle motion[19]. This can be explained based on the OAM density distribution, as shown in Fig. 7(c). The greater the intensity, the greater the OAM density, and the greater the acceleration when the particle moves there. The OAM density distributions along the z-axis in spatial space can be calculated with[20-22]

$$j_z = (r \times \varepsilon_0 \langle E \times B \rangle)_z = xS_y - yS_x \quad (6)$$

where $r = \sqrt{x^2 + y^2}$, **E** denotes an electric field and **B** denotes a magnitude field. $S_x$ and $S_y$ are the components along the $x$ and $y$ axes of the Poynting vector $S = \varepsilon_0 \langle E \times B \rangle$, respectively.

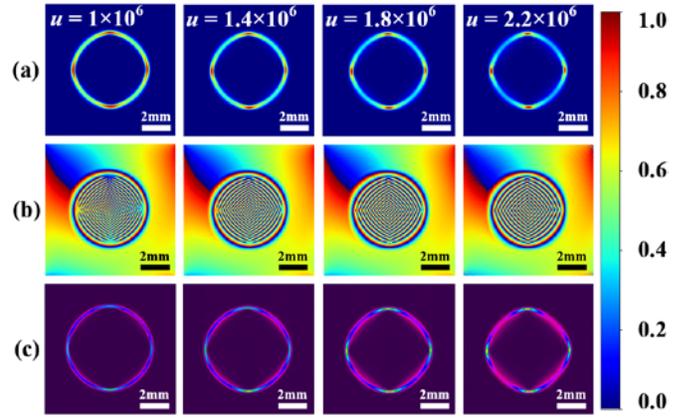

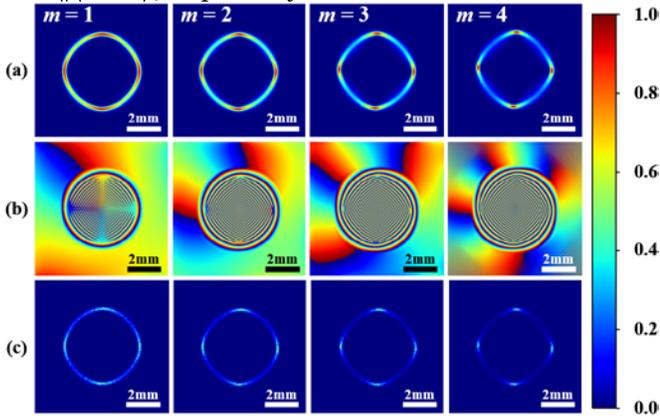

Fig. 7 The influence of TCs on the distribution of optical field, where (a) and (c) have been normalized. The order of the HOCP is 4. (a) Simulated intensity distributions of POVs modulated by the HOCP. (b) Simulated phase distributions of the HOCP corresponding to (a). (c) OAM density distributions of the HOCP corresponding to (a).

However, once the POV is determined, the TCs are settled. Is there a way to implement energy control similar to Fig. 7 for a determined POV? The answer is yes. Combined with the characteristics of a POV, we propose a method of using the HOCP to control the energy distribution of a POV. By altering the parameter $u$ from $1 \times 10$ to $2.2 \times 10^6$ of the POV, we can achieve the secondary energy control of the POV that has been generated, as shown in Fig. 8(a). With the gradual increase of the parameter $u$, the modulation effect on the already produced POV is similar to that in Fig. 7, and the experimental results are in good agreement, as shown in Fig. 8(c).

Fig. 8 Control the energy distribution of a POV via the HOCP, where (a) and (c) have been normalized. The order of the HOCP is 4. (a) Simulated intensity distributions of POVs modulated by the HOCP. (b) Simulated phase distributions of the HOCP corresponding to (a). (c) Experimental intensity distributions of the HOCP corresponding to (a).

Although the distribution of the optical field can be further controlled in both cases, the method we propose can maintain the mode purity of the original POV as much as possible while adjusting the shape of the optical field. The OAM spectrum for an arbitrary field $\Psi$ can be calculated with[23, 24]

$$P_m = \frac{C_m}{\sum_{-\infty}^{+\infty} C_n} \quad (7)$$

where

$$C_m = \int_0^\infty \langle a_m(r,\phi,z)^* | a_m(r,\phi,z) \rangle r dr \quad (8)$$

$$a_m(r,z) = 1/(2\pi)^{1/2} \int_0^{2\pi} \Psi(r,\phi,z) \exp(-im\phi) d\phi \quad (9)$$

and $(r,\phi,z)$ denotes the polar coordinate thereof. For simplicity and without loss of generality, the parameter $n$ ranges from -10 to +10. We calculate the mode purity according to the parameters used in Fig. 7 and Fig. 8, respectively. The calculated results are compared in Fig. 9. Although the effects of modulation are similar, the method of secondary energy control using HOCP has a much smaller impact on the model purity than altering TCs. The method we propose further extends the dimension of the POV modulation on the basis of ensuring the mode purity and provides more flexibility in optical micro-manipulation.

This can be explained in two ways. On the one hand, the CP causes astigmatism in a POV, and the larger the TCs, the greater the degree of astigmatism; on the other hand, the singularity of higher-order optical vortex tends to split towards lower-order singularities when encountering disturbances. The instability of the singularity caused a rapid decrease in the mode purity of the high-order POV, while astigmatism controlled only by the parameter $u$ retained the mode purity because of the stability of the low-order POV.

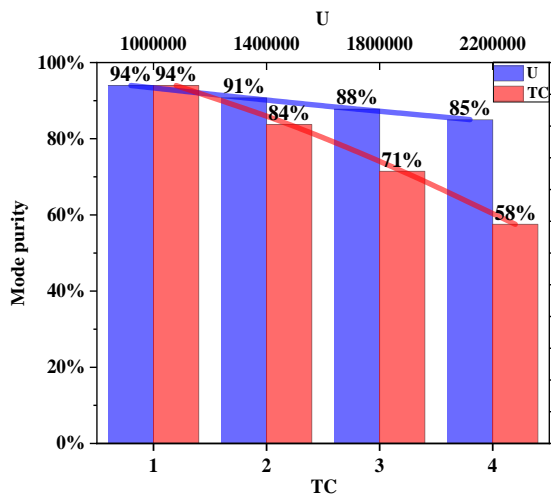

Fig. 9 The influence of parameter selection on model purity. As the TC increases from 1 to 4, the mode purity gradually decreases from 94% to 58%; as the parameter $u$ increases from $1\times10^6$ to $2.2\times10^6$, the mode purity gradually decreases from 94% to 85%.

In summary, we apply the CP to POVs for the first time experimentally. Firstly, we propose a method to measure the TCs +3 and ±5 of POVs with the LOCP at the Fourier plane. Compared with the far-field, this method can achieve the measurement of TCs in a short distance, saving space and optical components. Furthermore, this method realizes the in-situ measurement of POVs at the Fourier plane, which is meaningful in optical micro-manipulation. Secondly, we shape the intensity distributions of POVs to the triangle, quadrangle, and pentagon with the HOCP at the Fourier plane, respectively. On this basis, we show the asymmetric modulation ability of the HOCP for an optical field. The asymmetric modulation of the optical field greatly expands the range of the modulation of POVs and provides more flexibility in optical micro-manipulation. Finally, we discuss the influence of parameters on the distribution of the optical field. In this part, we propose a secondary energy control method that combines the characteristics of the POV, which ensures flexible control of the shape while reducing the impact on the orbital angular momentum of the original POV, compared with altering TCs.

This work was supported in part by the National Nature Science Foundation of China under Grant 11772001 and 61805283, and in part by the Key research projects of the foundation strengthening program and Outstanding Youth Science Foundation.